\documentclass[prb,twocolumn,secnumarabic]{revtex4}
\usepackage{amssymb}
\usepackage{amsmath}
\usepackage{graphicx}
\usepackage[sort&compress]{natbib}

\begin{document}

\title{Micellization of Sliding Polymer Surfactants}
\author{Vladimir A. Baulin$^1$, N.-K. Lee$^2$, Albert Johner$^{1,3}$ and
Carlos M. Marques$^{1,}$}
\email{marques@ics.u-strasbg.fr}
\affiliation{$^{1}$Institut Charles Sadron, CNRS UPR 22, 67083 Strasbourg Cedex, France. %
\vskip0.1cm $^{2}$Department of Physics, Sejong University, Seoul 143-743,
South Korea.\vskip0.1cm $^{3}$also LEA ICS/MPlP, Ackermannweg 10, 55128
Mainz, Germany.}

\begin{abstract}
Following up a recent paper on grafted sliding polymer layers (\textit{%
Macromolecules} \textbf{2005}, \textit{38}, 1434-1441), we investigated the
influence of the sliding degree of freedom on the self-assembly of sliding
polymeric surfactants that can be obtained by complexation of polymers with
cyclodextrins. In contrast to the micelles of quenched block copolymer
surfactants, the free energy of micelles of sliding surfactants can have two
minima: the first corresponding to small micelles with symmetric arm
lengths, and the second corresponding to large micelles with asymmetric arm
lengths. The relative sizes and concentrations of small and large micelles
in the solution depend on the molecular parameters of the system. The
appearance of small micelles drastically reduces the kinetic barrier
signifying the fast formation of equilibrium micelles.
\end{abstract}

\maketitle

\section{Introduction}

Predicting, controlling and finely tuning the self-assembly
properties of amphiphiles through molecular design is a problem of
central importance in physical chemistry.\cite{Safran,Israelash} It
is arguably also one of its major contributions to other fields: in
the biological realm, where self-assembled phospholipids build the
walls of liposomes and cells; in cosmetics, pharmaceutics or
detergency, where many formulations are self-assembled solutions of
surfactants, phospholipids and other amphiphile molecules. In this
context, diblock copolymers have emerged as a paradigm for
self-assembly.\cite{Reiss,Hamley,Alexandrid,Eisenberg} Typically,
the insoluble block drives the chains to self-associate and the
compositional asymmetry between the soluble and insoluble blocks
defines the assembling geometry.\cite{Alexandrid} The many
possibilities for architecture building offered by polymer synthesis
and the development of polymer theory led to an unprecedented power
of molecular control over the self-assembling structures of these so
called macrosurfactants. It is nowadays possible, for instance, to
build different self-assembled structures from diblock, triblocks
and many other block copolymers, to introduce charges at different
places along the chains with single charge accuracy, to form
reversible or frozen structures, to combine flexible and
semi-flexible blocks, \textit{etc}. However, a major difficulty
still hinders progresses in the predictive power of micellization
theories. Indeed, the macromolecular character of such surfactants
implies that there is a large kinetic barrier for micelle formation,
and the thermodynamic predictions for typical quantities such as the
critical micellar concentration or the aggregation number are in
many cases only marginally relevant. In this paper we discuss a
novel macrosurfactant architecture, based on rotaxane inclusion
complexes,\cite{Nakashima} that can lead to a significant decrease
of the kinetic barrier to micellization.

\begin{figure}[t]
\begin{center}
\includegraphics[width=5cm]{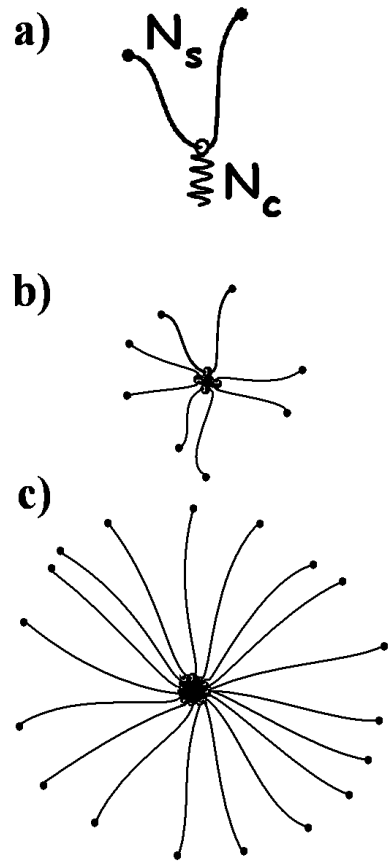}
\end{center}
\caption{Schematic representation of sliding a) unimers comprised of
soluble insoluble blocks of lengths $N_{s}$ and $N_{c}$,
respectively, b) small symmetric micelles and c) big asymmetric
micelles.} \label{slidmic}
\end{figure}

Rotaxanes are molecular complexes formed when a ring like molecule, the
rotor, is threaded over a linear molecule, the rotating axis.\cite{Nakashima}
Unthreading of the ring can be prevented by subsequent capping of the axis
ends.\cite{Ogino,OginoOhata} Rotaxanes can be made by combining different
linear polymers\cite{HaradaCoo,Wei} with different cyclic molecules, in
several solvents.\cite{Rekharsky} One of the most well studied systems
involve poly(ethylene-oxide) and $\alpha $-cyclodextrins, which are
oligosaccharides of $6$ glucose units assembled as rings. When an inclusion
complex formed by one $\alpha $-cyclodextrin and one PEO chain is attached
to a surface by grafting the $\alpha $-CD, it results a novel structure of
polymer layers for which we recently\cite{Baulin} coined the acronym SGP
layers, for sliding grafted polymer layers. Contrary to the usual
end-grafted polymer layer, in the SGP layers the chains retain the ability
to slide through the grafting ring, a new degree of freedom that allows, as
we recently have shown, to better relax the layer structure. In particular,
our work suggests that in spherical, star-like layers, as obtained for
instance if one attaches the $\alpha $-CD ring to a small colloid, sliding
of the polymers through the ring entails a truly versatile polymer layer
structure. At low number of grafts the chains adopt mostly symmetric
configurations with two comparable arm-sizes whereas with many grafts the
adopted configurations are asymmetric with essentially one long arm
participating in the corona. It was also shown that there exists an
intermediate range of grafting numbers where the number of arms
participating in the corona is constant and where the addition of more
grafts leads to the progressive replacement of symmetric configurations by
twice as much asymmetric ones. Conversely, in flat dense SGP layers, chains
adopt only asymmetric configurations, having hence the same structure as the
quenched ones. An important consequence for self assembly is that the free
energy of the curved SGP layers has a different structure from the usual
polymer layers grafted to spherical surfaces, but that the free energy of
flat SGP layers only marginally differs from their non-sliding equivalents.

We consider the micellization of decorated ``one-pearl necklaces'' where the
driving force for self-assembly is provided by the bead decoration, an
insoluble chain chemically attached to the ring molecule (see Figure \ref%
{slidmic}). The consequences of the sliding degree of freedom on the
self-assembly of sliding polymeric surfactants are discussed by focusing on
curved self-assembled structures, since for flat self-assembled layers only
marginal differences are expected with the usual grafted layers. After a
short reminder on diblock-copolymer micellization, we recall the theoretical
description of the spherical SGP layer, before a detailed discussion of the
possible micellization scenarios for this new family of macrosurfactant
architectures (Figure \ref{slidmic}).

\section{Micellization of diblock-copolymers\label{norm}}

In this section we briefly discuss the theory of micellization of block
copolymers, see \textit{e.g.} Ref. \onlinecite{Sens}. Consider a solution of
monodisperse block copolymer with $N_{c}$ insoluble and $N_{s}$ soluble
units. The insoluble blocks tend to aggregate and favor large aggregates
while coronas of soluble blocks oppose the formation of big micelles. The
free energy per unit volume $\digamma $, of the solution of non-interacting
micelles is a sum running over the aggregation number $p$:
\begin{equation}
\digamma =\sum\limits_{p=1}^{\infty }c_{p}\left[ k_{B}T\ln \left( \frac{c_{p}%
}{e}N_{T}b^{3}\right) +F_{p}\right]
\end{equation}%
where $F_{p}$ and $c_{p}$ is the free energy and the number density of a
micelle comprising $p$ surfactant chains respectively. The multiplier $%
N_{T}b^{3}$ with $N_{T}=N_{c}+N_{s}$ being the total volume of a copolymer.%
\cite{Nyrkova} This multiplier was often overlooked, a general discussion is
provided by Riess,\cite{Howard} who gives the relevant elementary volume, in
our case $N_{T}b^{3}$, for microemulsions, assemblies of droplets, etc. The
equilibrium distribution of chains in micelles is obtained by minimizing $%
\digamma $ with respect to $c_{p}$ along with the mass conservation
constraint:
\begin{equation}
\sum_{p=1}^{\infty }pc_{p}=\phi  \label{phi}
\end{equation}%
where $\phi $ is the total number of chains per unit volume. This gives the
equilibrium densities of $p$-arm micelles in the form $c_{p}b^{3}N_{T}=\exp
\left\{ (\mu p-F_{p})/k_{B}T\right\} $, where:
\begin{equation}
\mu =F_{1}+k_{B}T\ln (c_{1}N_{T}b^{3})  \label{mu}
\end{equation}%
is the chemical potential of unimers.

\begin{figure}[t]
\begin{center}
\includegraphics[width=8cm]{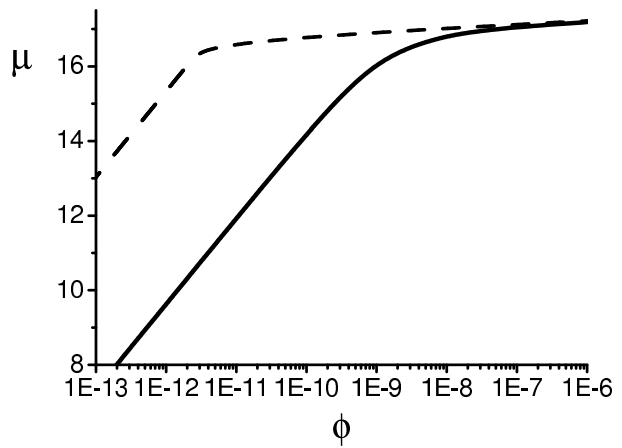}
\end{center}
\caption{Chemical potential $\mu $ as a function of total
concentration for quenched (dashed line) and sliding (solid line)
micelles. Parameters used: $N_{S}=300$, $N_{c}=62$ and $\sigma
=0.5$.} \label{mus}
\end{figure}

Thus we can rewrite the equilibrium $c_{p}$ as
\begin{equation}
c_{p}=\frac{\left( c_{1}N_{T}b^{3}\right) ^{p}}{N_{T}b^{3}}\exp \left( -%
\frac{F_{p}-pF_{1}}{k_{B}T}\right)  \label{distribution}
\end{equation}%
Rather than the distribution $c_{p}$, we will often use the more convenient
function $\Omega _{p}=\ln (c_{1}/c_{p})$:
\begin{equation}
\Omega _{p}=F_{p}-pF_{1}-\left( p-1\right) k_{B}T\ln \left(
c_{1}N_{T}b^{3}\right)  \label{Omega}
\end{equation}%
which may be considered as the thermodynamic potential. Given a value of $%
c_{1}$ in eq. \ref{distribution} allows us to calculate the whole
distribution $c_{p}$ and the corresponding total chain concentration $\phi $%
. Experimentally, the chemical potential $\mu $ is often studied as a
function of $\phi $. We obtained the chemical potential curve from eqs.(\ref%
{phi})-(\ref{distribution}) with $c_{1}$ as the parameter. A kink in the
representation (Figure \ref{mus}, dashed line) is commonly used as an
indicator for the onset of micellization. Usually only a narrow range of
unimer concentrations at equilibrium $c_{1}$ falls in the physically
acceptable condition: $N_{T}\phi b^{3}\ll 1$. When no interaction between
micelles is accounted for, the more stringent criterion $%
\sum_{p}c_{p}V_{p}<1 $ should be obeyed where $V_{p}$ is the volume of a
micelle comprising $p$ surfactants.

The micellization scenario depends on the precise form of the free energy $%
F_{p}$, discussed in the following paragraphs for quenched copolymers and
annealed sliding copolymers.

\section{Free energy of star-like micelles}

The free energy of a micelle contains the molecular characteristics of the
micellization process. Thus, in order to investigate the micellization of
sliding polymer surfactants, we have to specify the explicit form for the
corresponding $F_{p}$. The free energy of a micelle is the sum of the core
and of the corona contributions. Hereafter energies are expressed in $k_{B}T$
units.

\subsection{Core contribution}

We assume that the insoluble blocks form a dense homogeneous core where
soluble blocks and the solvent cannot penetrate. Hence the micelle free
energy has two core contributions: (i) the surface tension term, $F_{c}=4\pi
\sigma R_{c}^{2}$, where $R_{c}$ is the radius of the core and $\sigma $ is
the core-solvent surface tension expressed in $k_{B}T/b^{2}$ units. For an
incompressible core of size $R_{c}=(3/(4\pi )pN_{c})^{1/3}b$, this leads to $%
F_{c}=c\sigma N_{c}^{2/3}p^{2/3}$, where $c=(36\pi )^{1/3}$. (ii) Gaussian
elastic contributions arising from stretching of the insoluble blocks in the
core, $F_{el}=wpR_{c}^{2}/(N_{c}a^{2})=wp^{5/3}/N_{c}^{1/3}$, where $w=3\pi
^{2}/80$ reflects the spherical geometry of the core. The value of $w$ is
obtained from SCF theory in the strong stretching limit.\cite{Zhulina}
Although the elastic contribution of the core is only larger than $k_{B}T$
for micelles with large cores, we keep it for convenience.\cite{foot1}

\subsection{Corona contribution}

For large soluble blocks, as considered here, the corona is usually
envisioned as a star of soluble blocks radially stretched away from the
spherical core. The partition function $Z_{p}$ of a star with $p$ equal arms
of contour length $N_{s}$ is given by the critical exponent $\gamma _{p}$, $%
Z_{p}\sim N_{s}^{\gamma _{p}-1}$.\cite{Duplantier} Star exponents $\gamma
_{p}$ are known exactly in two dimensions and for ideal chains ($d\geq 4$).
In three dimensions $\epsilon $ -expansions ($\epsilon =4-d$) are available,
to first order:\cite{Duplantier}

\begin{equation}
\gamma _{p}-1=-{\frac{\epsilon }{16}}p(p-3)+o(\epsilon ^{2})  \label{epsi}
\end{equation}%
Recently Monte Carlo simulations were carried out by Hsu \textit{et al.}\cite%
{Grassberger} in order to find the exact values of $\gamma _{p}$ for a large
range of $p$ values. The results do not quantitatively agree with the
classical predictions of the Daoud-Cotton model\cite{Daoud} $\gamma
_{p}-1\sim -p^{3/2}+\ldots $, indicating that the asymptotic limit of very
large $p$ numbers, where subdominant powers of $p$ are negligible, is not
yet reached. If we insist on fitting Hsu \textit{et. al.} results to the
Daoud and Cotton model, say between $p=20$ and $p=60$, where the simulated
arms are still fairly long, the obtained amplitude is close to $0.2$. It
seems that the asymptotic limit, restricted to the leading term, is also out
of experimental reach. Authors showed that the best fit of the simulation
data with a power law $\sim -p^{z}$ is obtained with $z=1.68$. Our
discussion below is based on the simulation data and we will use, for
convenience, the fitting function of the form close to eq. (\ref{epsi}): $%
-p(bp-a)^{z}/16$ reflecting the simulation results. The best fit is

\begin{equation}
\gamma _{p}-1=-\frac{p}{16}\left( 1.5p-6\right) ^{0.7},\text{ }p>4
\label{gam}
\end{equation}%
while for $p<4$ we take values from the table of Ref. %
\onlinecite{Grassberger}.

When there is sliding degree of freedom, the free energy of the corona has a
more complex structure. The possibility for the soluble blocks to slide
through a ring results in an annealed arm length distribution in the coronas
of such micelles made of $p$-chains. Adjusting the length of the arm in the
corona can decrease the crowding effect originating from the steric
repulsions between soluble blocks. In our previous paper\cite{Baulin} we
found three possible regimes for the sliding coronas depending on the number
of sliding chains per aggregate. The transition between these regimes is
determined from a threshold value of the number of arms $p^{\ast }$, defined
by the conditions $\gamma _{p^{\ast }}-\gamma _{p^{\ast }-1}>-1$ and $\gamma
_{p^{\ast }+1}-\gamma _{p^{\ast }}\leq -1$. If the number of chains per
aggregate is small, $p\leq p^{\star }/2$, the corona is fully annealed and
the sliding chains are likely to adopt any configuration. We call such
micelles symmetric micelles, since symmetric configurations are prevalent.
The corona properties are determined by configurations with $2p$ arms and
its partition function is given by\cite{Baulin}

\begin{equation}
Z_{p}\propto N_{s}^{p}N_{s}^{\gamma _{2p}-1},\text{ }p\leq p^{\star }/2
\label{smallp}
\end{equation}%
For a larger number of grafting chains, in the intermediate regime, $%
p^{\star }/2<p<p^{\ast }$, only $p^{\star }-p$ chains adopt symmetric
configurations, while the rest of the chains are stretched out from the
ring. The corona has $p^{\ast }$ arms and its partition function reads

\begin{equation}
Z_{p}\propto N_{s}^{p^{\star }-p}N_{s}^{\gamma _{p^{\star }}-1},\text{ }%
p^{\star }/2<p<p^{\ast }  \label{intermediatep}
\end{equation}%
For even larger micelles, $p\geq p^{\star }$, all the chains in the corona
are strongly asymmetric and the partition function of the corona coincide
with the partition function of a quenched star:

\begin{equation}
Z_{p}\propto N_{s}^{\gamma _{p}-1},\text{ }p\geq p^{\star }  \label{largep}
\end{equation}

\begin{figure}[t]
\begin{center}
\includegraphics[width=7.5cm]{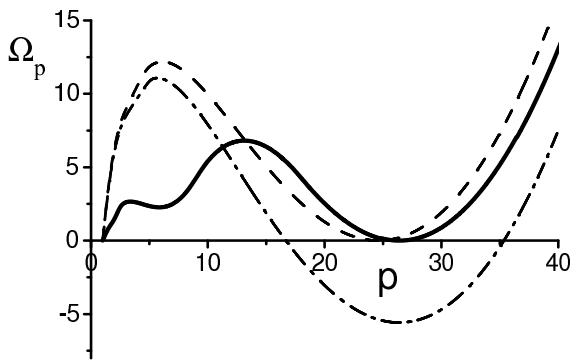}
\end{center}
\caption{Potential $\Omega _{p}=\ln (c_{1}/c_{p})$ as a function of
the aggregation number $p$ for sliding micelles (solid line, total
surfactant concentration $\phi =3.\times 10^{-8}$) and for quenched
micelles (dashed line, $\phi =1.\times 10^{-10}$, dash-dotted line,
$\phi =3.\times 10^{-8}$, same as for sliding micelles) for the set
of parameters: $N_{S}=300$, $N_{c}=70$ and $\sigma =0.5$. Micelles
form almost without kinetic barrier from sliding polymer
surfactants.} \label{Omegas}
\end{figure}

The determination of the threshold value of the number of arms $p^{\ast }$
strongly depends on our knowledge of the values of the critical exponents $%
\gamma _{p}$. Our previous estimates\cite{Baulin} were based on the first
order $\epsilon $-expansion (Eq. \ref{epsi}) yielding $p^{\ast }=9$.
However, if we use the values obtained from the computer simulations in ref. %
\onlinecite{Grassberger}, we get $p^{\ast }=18$. Thus, in the following we
assume the latter estimate as most accurate. We use our fitting function (%
\ref{gam}) for the values of $\gamma _{p}$, and the free energy of the
sliding corona of a micelle $F_{p}^{corona}=-\ln Z_{p}$ is determined as a
piecewise function of the number of chains $p$.

\subsection{Total free energy}

Combining all terms together, the energy of a micelle is given by

\begin{equation}
F_{p}=F_{p}^{corona}+c\sigma N_{c}^{2/3}p^{2/3}+wp^{5/3}/N_{c}^{1/3}
\end{equation}%
with
\begin{equation}
F_{p}^{corona}=f(p)\ln N_{s}  \label{Fquen}
\end{equation}%
where $f(p)$ is defined piecewise by eqs.(\ref{smallp},\ref{intermediatep},%
\ref{largep}) as discussed in the previous section for sliding copolymers
whereas for quenched copolymers: $f(p)=-(\gamma _{p}-1)$.

Note that the corona free energy of sliding surfactant unimers ($p=1$) is $%
F_{1}^{s}=-\gamma _{1}\ln N_{s}$ while that of the quenched unimer is $%
F_{1}^{q}=-(\gamma _{1}-1)\ln N_{s}$. Thus the free energy of
sliding surfactant is about $-\ln N_{s}$ less than that of quenched
ones with the same soluble block size.

Next we use the expressions of the free energy $F_p$ to compute the micelle
size distribution for both quenched and sliding copolymers.

\section{Micelle-size-distribution: Sliding vs. Quenched copolymers\label%
{slid}}

A preferred aggregation number is reflected by a minimum in the potential $%
\Omega _{p}$. Roughly speaking, the corresponding aggregates will dominate
over unimers if the minimum potential value is negative. For usual quenched
copolymers the potential has only one, pretty sharp, minimum (see Figure \ref%
{Omegas}) and the polydispersity of spherical micelles is usually very low,
the distribution $c_{p}$ being sharply peaked around the average aggregation
number $p_{m}$. In this sense, quenched copolymer association should be
termed $\emph{closed}$ association where the distribution is dominated by
aggregates of uniform size with weak fluctuations appearing at a well
defined unimer concentration. In the common used classification, $\emph{open}
$ association means that aggregates of different sizes significantly
contribute to the distribution and appear gradually. The practical relevance
of this classification has been critically discussed recently.\cite{Nyrkova2}

The average aggregation number and the \textit{CMC} are found from the
conditions: $\Omega _{p}=\partial \Omega _{p}/\partial p=0$.\cite{foot2} The
minimum is separated from the origin by a kinetic barrier. Thus it takes a
typical time $\sim \exp (-U_{\max }/k_{B}T)$ for isolated chains to form a
micelle; with $U_{\max }$ the maximum of the barrier. A rough scaling
estimate gives $U_{\max }/kT\sim \sigma N_{c}^{6/5}$.\cite{foot3} Depending
on the conditions, this time for usual block copolymer micelles can be very
large. Very often, quenched copolymer micelles do not form over reasonable
time scales at the \textit{CMC}, where they are thermodynamically favored.
Only at much higher concentrations, the barrier $U_{\max }/kT$ becomes low
(see Figure \ref{Omegas}).

In contrast to the quenched case, the potential $\Omega _{p}=\ln
(c_{1}/c_{p})$ (Eq. (\ref{Omega})) for sliding micelles can have two minima.
One corresponds to small symmetric micelles with aggregation numbers up to $%
9 $, and the second corresponds to big asymmetric micelles (Figure \ref%
{Omegas}). The appearance of small micelles leads to a drastic decrease of
the kinetic barrier, such that the first minimum is separated from the
origin by a barrier of order $kT$ signifying the fast formation of
equilibrium micelles. The chemical potential $\mu $ of sliding copolymers as
a function of $\phi $ presents a rounded kink as compared to the quenched
case (Figure \ref{mus}).

\begin{figure}[t]
\begin{center}
\includegraphics[width=8.5cm]{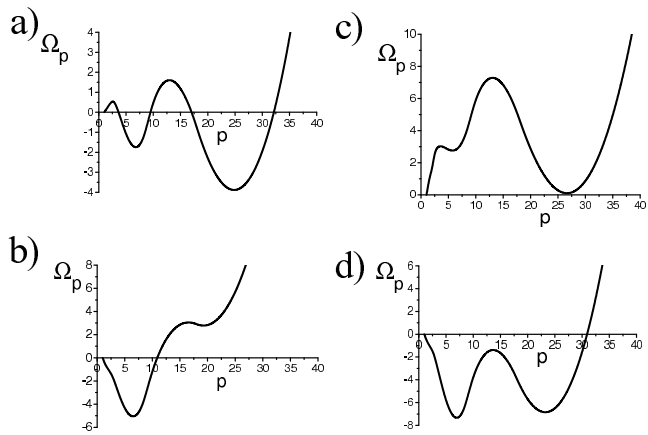}
\end{center}
\caption{Typical examples of the potential $\Omega
_{p}=\ln (c_{1}/c_{p})$ for different set of parameters: a) $N_{S}=300$, $%
N_{c}=60$, $\sigma =0.5$, $\phi =3.3\times 10^{-5}$, b) $N_{S}=300$, $%
N_{c}=55$, $\sigma =0.6$, $\phi =2.0\times 10^{-8}$, c) $N_{S}=1000$, $%
N_{c}=60$, $\sigma =0.5$, $\phi =2.8\times 10^{-5}$, d) $N_{S}=10000$, $%
N_{c}=60$, $\sigma =0.7$, $\phi =7.8\times 10^{-7}$. Depending on
parameters, small micelles appear first, a) and c), or formation of
big micelles happens earlier, b) and d).} \label{example}
\end{figure}

\begin{figure}[t]
\begin{center}
\includegraphics[width=4cm]{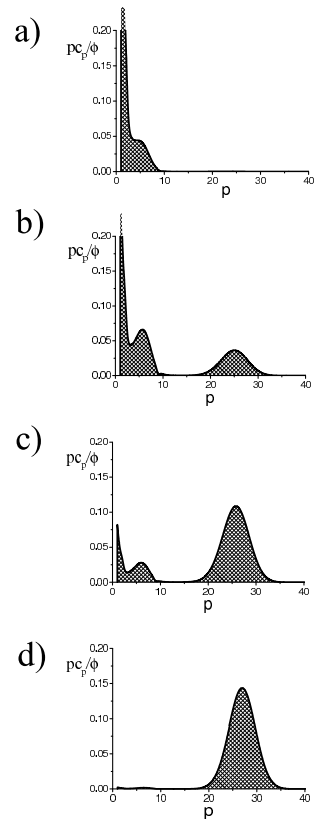}
\end{center}
\caption{Evolution of the mass distribution in micelles, $%
pC_{p}/\phi $, with increasing total concentration $\phi $: a)
$2.\times 10^{-10}$, b) $5.\times 10^{-10}$, c) $2.\times 10^{-9}$,
d) $8.\times 10^{-8}$ for $N_{S}=300$, $N_{c}=70$ and $\sigma
=0.5$.} \label{Cp}
\end{figure}

\begin{figure}[t]
\begin{center}
\includegraphics[width=8cm]{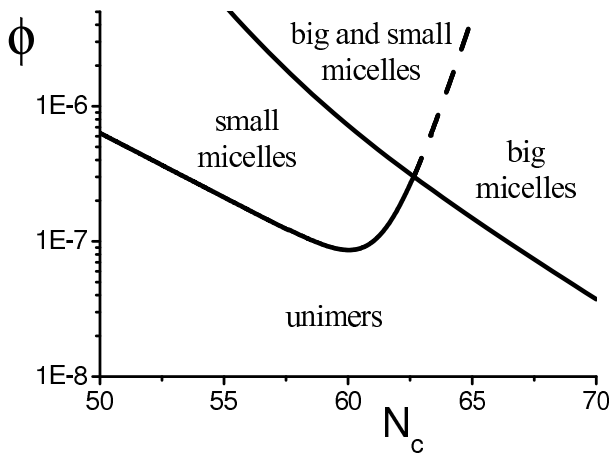}
\end{center}
\caption{Schematic diagram of different types of micelles present in
a solution. Labels designate regions where the indicated structures
are dominant. Parameters used: $N_{S}=300$, $\sigma =0.5$. The
boundary between the region of big micelles and the region of the
coexistence of small and big micelles is indicated as dashed line,
because big micelles usually dominate by mass in both regions.}
\label{diagram}
\end{figure}

\begin{figure}[t]
\begin{center}
\includegraphics[width=8.5cm]{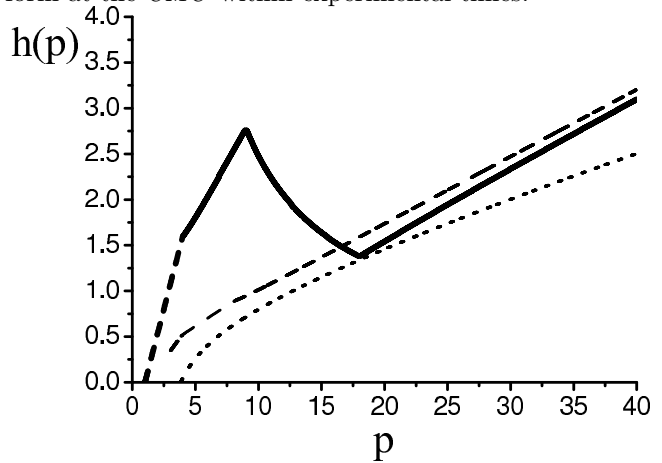}
\end{center}
\caption{Graphical solution of the eq. (\ref{apprpm}). For values of
$\beta =\frac{1}{3}c\sigma N_{c}^{2/3}/\ln N_{s}$ between $1.38$ and
$2.77$ a small symmetric micelle and a large asymmetric one can be
more stable with respect to the unimer (for different chemical
potentials). The bold dashed line ($p<3$) was not calculated. The
thin dashed line corresponds to quenched copolymers, where there is
only one micelle size. The dotted line corresponds to the
Daoud-Cotton model for quenched copolymers.} \label{X}
\end{figure}

Depending on molecular parameters of the system, the formation of small
micelles can follow (Figure \ref{example}a and \ref{example}c) or precede
(Figure \ref{example}b and \ref{example}d) the formation of big micelles.
Typical examples of the mass distribution in micelles, $pc_{p}/\phi $, with
increasing polymer concentration are shown in Figure \ref{Cp}. The
equilibrium aggregation number switches between symmetric (small) and
asymmetric (big) micelles as the concentration increases. Tuning the
parameters of the system, either the surface tension of the core, $\sigma $,
or the relative length of the blocks (in our case we vary $N_{c}$ keeping $%
N_{s}$ constant), we can change the order of the appearance of small and big
micelles. Figure \ref{diagram} shows the corresponding state diagram which
is related to Figure \ref{X}. When big and small micelles coexist the big
ones usually dominate by mass. Hence we represent the boundary between big
and big+small micelles by a dashed line.

Though the full potential (Eq.\ref{Omega}) is uneasy to handle analytically,
we may somewhat simplify it by omitting the stretching of the insoluble
blocks. The aggregation number of a vanishing minimum of this approximate
potential (for some chemical potential) is then solution of the following
equation:
\begin{equation}
\frac{(p-1)\frac{\partial f(p)}{\partial p}-f(p)+f(1)}{p^{2/3}+2p^{-1/3}-3}={%
\frac{1}{3}}\frac{c\sigma N_{c}^{2/3}}{\ln N_{s}}  \label{apprpm}
\end{equation}%
the left hand side of this equation $h(p)$ collects all the terms depending
on $p$. Thus, in this approximation the desired aggregation numbers depend
only on the combination of parameters $\beta ={\frac{1}{3}}c\sigma
N_{c}^{2/3}/\ln N_{s}$. It is easy to see that in the quenched copolymer
case there is indeed only one solution of $p_{m}$ and in the Daoud-Cotton
limit $p_{m}\sim \left( \sigma N_{c}^{2/3}/\ln N_{s}\right) ^{6/5}$, a
classical result.

It is worthwhile discussing the number of solutions of eq. (\ref{apprpm}) as
a function of the parameter $\beta $ for sliding copolymers. It can be
obtained by counting the number of intersections of $h(p)$ (Figure \ref{X})
with the line parallel to the abscissa. In the first region, $\beta \lesssim
1.38$, there is only one intersection. It corresponds to one minimum in $%
\Omega _{p}(p)$. The number of arms $p$ is small, thus this region
corresponds to the coexistence of unimers with small symmetric micelles. In
the region $1.38\lesssim \beta \lesssim 2.77$ there are three intersections
showing the existence of two minima separated by a barrier. Small micelles
with aggregation numbers $p<p^{\ast }/2=9$ coexist with big micelles with $%
p>p^{\ast }=18$. In the third regime, $\beta \gtrsim 2.77$, there are only
asymmetric micelles of high aggregation numbers. The crossover values of the
parameter $\beta $ are compatible with Figure \ref{diagram}, where the
crossover values of $N_{c}$ are $40$ and $93$. For given values of
parameters ($N_{c},$ $N_{s},$ $\sigma $) it should further be checked that
the micelles exist for physical concentrations.

Figure \ref{X} also shows the aggregation number of quenched
copolymer micelles as a function of $\beta$ (dashed line). The
difference with large sliding micelles is due to a shift in the free
energy of a unimer from $F_{1}^{s}$ to $F_{1}^{q}$. The Daoud-Cotton
line is also shown (dotted line). As already discussed, no agreement
can be obtained using a simple Daoud-Cotton power law.

\section{Concluding remarks\label{secconclusion}}

Sliding-diblock-copolymer surfactants where the soluble and insoluble blocks
are topologically tethered by a small ring polymer show a much reacher
micellization behavior than the corresponding quenched diblock copolymers.
When the size of insoluble blocks is not too long, both small micelles (with
an aggregation number slightly smaller than $p^{\ast }/2=9$) and large
micelles (with an aggregation number slightly larger than $p^{\ast }=18$)
may coexist in the solution. In the small micelle each copolymer
participates in the corona with two arms whereas in the large micelle the
asymmetric arm length distribution is dominant. Hence coronas in small and
large micelles comprise a similar number of arms. Small micelles can form
without appreciable kinetic barrier at the lowest concentration where they
are thermodynamically favored. The barrier opposing the formation of the
larger micelle remains modest. This is in marked contrast to the
corresponding quenched copolymers where kinetic barriers are very high at
the same concentrations. Which micelle appears first with increasing
concentration depends on the block asymmetry. At somewhat higher
concentrations, the micelle-size-distribution broadens and covers the whole
range from the small to the large aggregation number described previously.

For large insoluble blocks only the large micelles, similar to the quenched
ones, form. Due to the loss of the sliding degree of freedom upon
association, the annealed \textit{CMC} is higher than the quenched one. On
the other hand the kinetic barrier is markedly lower for annealed copolymer
micelles (typically by a factor 2) that may form at the \textit{CMC} within
experimental times.

Our description of star-like micelles is based on excellent values of the
vertex exponents from a recent simulation by Hsu \textit{et. al.}\cite%
{Grassberger} We think that this improves over the standard Daoud and Cotton
model for the classic quenched micelles.

Sliding copolymers are a new interesting example of non ionic
an\-nealed copolymers. Some charged micelles can also display a
similar behavior as reported by Zhulina and Borisov for
insoluble/annealed-polyelectrolyte diblocks.\cite{Zhulinaannealed}
To our knowledge only flat brushes of complexing polymers have been
studied theoretically.\cite{Currie} Diblocks where the soluble block
forms a complex with small colloids (proteins) also belong to the
class of annealed copolymers. One may speculate whether such
diblocks also form two types of micelles.

\begin{acknowledgments}

V. B. gratefully acknowledges the French Space Agency, CNES for a
research post-doctoral fellowship. N. K. L. acknowledges financial
support from KOSEF/CNRS exchange program N18055.

\end{acknowledgments}


\begin{thebibliography}{99}
\bibitem{foot1} For unimers and small aggregates a small compression term
could rather be taken into account, which is neglected.
\bibitem{foot3} This is based on the fact that the stretching energy of the
corona and the surface tension of the core are comparable at the \textit{CMC}
and at the maximum of the barrier.
\end{thebibliography}
\end{document}